\documentclass[12pt,preprint]{aastex}

\newcommand{\etal}{et~al.\ }

\newcommand{\flux}{\hbox{erg~cm$^{-2}$~s$^{-1}$}}

\newcommand{\be}{\begin{equation}}
\newcommand{\ee}{\end{equation}}
\newcommand{\ba}{\begin{eqnarray}}
\newcommand{\ea}{\end{eqnarray}}

\newcommand{\chandra}{{\emph{Chandra}}}

\newcommand{\swift}{\emph{Swift}}

\begin{document}

\def\Mesz{M\'esz\'aros}
\def\sarc{$^{\prime\prime}\!\!.$}
\def\arcsec{$^{\prime\prime}$}
\def\ls{\lower 2pt \hbox{$\;\scriptscriptstyle \buildrel<\over\sim\;$}}
\def\gs{\lower 2pt \hbox{$\;\scriptscriptstyle \buildrel>\over\sim\;$}}

\title{Go Long, Go Deep: Finding Optical Jet Breaks for Swift-Era GRBs with the LBT\altaffilmark{1}}

\author{X.~Dai\altaffilmark{2}, P.~M.~Garnavich\altaffilmark{3}, J.~L.~Prieto\altaffilmark{2}, 
K.~Z.~Stanek\altaffilmark{2}, C.~S.~Kochanek\altaffilmark{2}, 
J.~Bechtold\altaffilmark{4}, N.~Bouche\altaffilmark{5}, P.~Buschkamp\altaffilmark{5},
E.~Diolaiti\altaffilmark{6}, X.~Fan\altaffilmark{4}, E.~Giallongo\altaffilmark{7}, 
R.~Gredel\altaffilmark{8}, J.~M.~Hill\altaffilmark{9}, 
L.~Jiang\altaffilmark{4}, C.~McClellend\altaffilmark{3},
P.~Milne\altaffilmark{4},  F.~Pedichini\altaffilmark{7}, R.~W.~Pogge\altaffilmark{2}, R.~Ragazzoni\altaffilmark{10}, J.~Rhoads\altaffilmark{11}, R.~Smareglia\altaffilmark{12},
D.~Thompson\altaffilmark{9}, R.~M.~Wagner\altaffilmark{2,9}
}

\altaffiltext{1}{Based on data acquired using the Large Binocular
Telescope
 (LBT).  The LBT is an international collaboration among
institutions in the United States, Italy and Germany. LBT Corporation
partners are: The University of Arizona on behalf of the Arizona
university system; Istituto Nazionale di Astrofisica, Italy; LBT
Beteiligungsgesellschaft, Germany, representing the Max-Planck Society,
the Astrophysical Institute Potsdam, and Heidelberg University; The Ohio
State University, and The Research Corporation, on behalf of The
University of Notre Dame, University of Minnesota and University of
Virginia.}  
\altaffiltext{2}{ Department of Astronomy, Ohio State University, Columbus, OH 43210, USA}
\altaffiltext{3}{ University of Notre Dame, 225 Nieuwland Science, Notre Dame, IN 46556-5670, USA}
\altaffiltext{4}{ Steward Observatory, The University of Arizona, Tucson, AZ 85721, USA}
\altaffiltext{5}{ Max Planck Institut fur extraterrestrische Physik, Giessenbachstrasse, D-85748 Garching, Germany}
\altaffiltext{6}{ INAF, Osservatorio Astronomico di Bologna, via Ranzani 1, I-40127 Bologna, Italy}
\altaffiltext{7}{ INAF, Osservatorio Astronomico di Roma, via di Frascati 33, I-00040
Monteporzio, Italy}
\altaffiltext{8}{ Max-Planck Institute for Astronomy, Konigstuhl 17, 69117 Heidelberg, Germany}
\altaffiltext{9}{ Large Binocular Telescope Observatory, University of Arizona, 933 N. Cherry Ave., Tucson, AZ  85721-0065, USA}
\altaffiltext{10}{ INAF, Osservatorio Astronomico di Padova, vicolo dell'Osservatorio 5, I-35122 Padova, Italy}
\altaffiltext{11}{ School of Earth and Space Exploration, Arizona State University, Tempe, AZ, USA}
\altaffiltext{12}{ INAF, Osservatorio Astronomico di Trieste, via G. B. Tiepolo 11, I-34131 Trieste, Italy}


\begin{abstract}
Using the 8.4m Large Binocular Telescope, we observed six GRB afterglows from 2.8 hours to 30.8 days after the burst triggers to systematically probe the 
late time behaviors of afterglows including jet breaks, flares, and supernova
bumps.  We detected five afterglows with Sloan $r$' magnitudes ranging from 23.0--26.3 mag. 
The depth of our observations allows us to extend the temporal baseline
for measuring jet breaks by another decade in time scale.
We detected two jet breaks and a third candidate, all of which are not detectable without deep, late time optical observations.  In the other three cases, we do not detect the jet breaks either because of contamination from the host galaxy light, the presence of a supernova bump, or the intrinsic faintness of the optical afterglow. 
 This suggests that the basic picture that GRBs are collimated is still valid and that the apparent lack of \swift\ jet breaks is due to poorly sampled
afterglow light curves, particularly at late times.
\end{abstract}


\section{Introduction}

It is a basic ingredient in GRB models that GRBs are collimated in jets (e.g., Rhoads 1999).
This assumption affects the estimates of both the total energy of GRBs and the rate of GRBs.
Depending on the GRB jet model, a beaming correction of $\sim20$--$500$ is needed to obtain the intrinsic rate of GRBs
(e.g., Frail et al. 2001; Granot et al. 2002; Zhang et al. 2004; Yamazaki et al. 2004; Dai \& Zhang 2005), and 
uncertainties in this correction
need to be resolved before we can fully understand GRBs and their progenitors.
Before the launch of \swift, jet breaks had been observed in many optical light curves (see Zeh et al.\ 2006 for a collection of pre-\swift\ jet breaks) and several bursts had achromatic breaks across the optical bands (e.g., GRB~990510, Stanek et al.\ 1999), consistent with the predictions of the jet model.
After the launch of \swift\ (Gehrels et al.\ 2004), with its rapid localization of GRBs and the dedicated on-board XRT instrument, the number of GRBs with jet breaks was expected to increase significantly.
Instead, the afterglow light curves were found to be more complicated, especially in the X-ray band, where multiple breaks or giant X-ray flares are present in more than half of the \swift-XRT light curves (Nousek et al.\ 2006; O'Brien et al.\ 2006).  The optical afterglow light curves also show complicated decay patterns both in pre-\swift\ bursts (e.g. Bersier et al. 2003) and \swift\ bursts (e.g., Stanek et al.\ 2007).  
With these complications in the afterglow light curves, it can be difficult to identify a jet break.
As a result, the number of jet breaks detected in the \swift\ era is rather small (e.g., GRB~050525A, Blustin et al.\ 2006, GRB~060206, Stanek et al.\ 2007, and GRB~060526, Dai et al.\ 2007, and see Liang et al.\ 2007 for a comprehensive analysis).  
Indeed, the absence of late time X-ray breaks can be interpreted as a
challenge to the basic beaming model (e.g., Burrows \& Racusin 2007a)
or the fireball model (Dado et al. 2007).

While the flares are complications, the biggest problem for identifying
jet breaks is the need for well-sampled, long term light curves.
While almost all \swift\ bursts have good signal-to-noise ratio (S/N) X-ray light curves at early decay times,
most of the jet breaks occur at late times when the uncertainties in the XRT light curves increase significantly.  In these cases, the claim that the data are consistent with the single power-law decay does not distinguish between models, because the error-bars are so large that the data are also consistent with the broken power-law of a jet break ---
jet breaks are ``hidden'' in low S/N 
light curves or ``masked'' by additional sources of emission (Curran et al. 2007; Ghirlanda et al. 2007; Sato et al. 2007).
Shao et al. (2007) have also proposed that jet breaks are hidden by X-ray light echos from dust scattering.
In the optical bands, the afterglow monitoring is generally sparse, in part 
because of the large number of bursts detected.  
However, the late-time afterglow can be measured accurately using large optical telescopes.
In fact, all the \swift\ bursts where jet breaks have been detected have well-sampled optical light curves (e.g., Dai et al.\ 2007).

Given the need for deep, late-time monitoring of GRBs, we initiated an observing program using the newly built 8.4 meter Large Binocular Telescope (LBT, Hill et al. 2006).  
With 20--30 minute exposures, we are able to reach a flux limit of $\sim25$--$26$~mag in the Sloan $r'$ band.  
This allows us to extend the temporal baselines of optical afterglow monitoring by roughly an order of magnitude, and systematically search for the jet breaks and other emission components in the optical light curves.

\section{Observations and Data Reduction\label{sec:obs}}
We observed 6 GRBs, GRB~070125 (Hurley et al. 2007), GRB~070311 (Mereghetti et al. 2007), GRB~070411 (Moretti et al. 2007), GRB070412 (Romano et al. 2007), 
GRB070419A (Stamatikos et al. 2007), 
GRB070518 (Guidorzi et al. 2007), with the 8.4~$m$ LBT (Hill et al. 2006) 
using the Large Binocular Camera (LBC) Blue CCD camera 
(Ragazzoni et al. 2006) during the LBT/LBC Science Demonstration Time period in
 January--November 2007.  
A journal of
the LBT observations is presented in Table~1.
We took 5--25 dithered, 200~s exposures for each epoch
using the SDSS $r$-band filter. 
All the
images from a given epoch were registered to a common astrometric
grid
and co-added. We ran DAOPHOT/ALLSTAR (Stetson 1992) on
the
deep, co-added images to obtain instrumental magnitudes of all the detected
point-sources. The magnitudes of the GRBs were calibrated relative to
bright stars, either a single star from the USNO-B
catalog or several
stars with SDSS photometric data.
We gathered other $R$ or $R_c$ bands data for these bursts from GCN circulars and the literature to complete the optical light curves. 
While that there are slight differences between $R$, $R_c$, and Sloan $r'$ filters, 
this has little consequence for our analysis.

We also reduced the \swift-XRT light curves for the sample.  
We started from the XRT level 2 event files for the photon counting (PC) mode, where events were filtered to be in the 0.2-10 keV energy band and restricted to grades 0--12.  
We extracted the XRT-PC spectra and light-curves with the software tool \verb+xselect+.
 We used the \verb+rmf+ files from the standard XRT calibration distribution, generated the \verb+arf+ files with the \emph{Swift}-XRT software tool 
and fit the X-ray spectra with \verb+XSPEC+ (Arnaud 1996).

\section{The Afterglow Light Curves\label{sec:lc}}

\subsection{GRB~070125}
GRB~070125\footnote{An independent analysis of GRB~070125 is provided by Updike et al.\ (2008)} ($z=1.547$, Fox et al.\ 2007) was relatively well observed in the optical bands, and GCN data sample between $6\times10^4$ and $2.3\times10^6$~s after the burst trigger.
The optical data from $9\times10^4$ to $4\times10^5$~s can be well fit by a power-law 
model (dashed line in Figure~1) with a slope of $\alpha=1.65\pm0.05$ ($\chi^2/dof = 7.4/14$).
The first LBT data point ($r' = 26.3\pm0.3$~mag at 26.8 days) deviates by $6.4\sigma$ from the single power-law fits.
The non-detection of the second LBT observation ($r' < 26.2$~mag at 298 days) indicates that the first
LBT detection is mostly from the afterglow emission with little host galaxy contamination.
The data suggests an optical break between the LBT observations and the early data taken before $4\times10^5$~s, 
which is also supported by the upper limit of $R > 23.8$~mag at $10^6$~s from MDM (Mirabal et al.\ 2007).
We fit the optical data with a broken power-law as $f(t) \propto ((t/t_b)^{\alpha_1s}+(t/t_b)^{\alpha_2s})^{-1/s}$ where we fixed the smoothness parameter to $s=2.5$, and obtained $\alpha_1=1.58\pm0.12$, $\alpha_2=2.87\pm0.17$, and $t_b = (5.0\pm1.6)\times10^5$~s with $\chi^2/dof = 6.7/13$.

GRB~070125 was also observed by \swift\ and \chandra.
We compared the X-ray light curve to the optical light curve, and found that they are broadly consistent with each other.
We fit the X-ray data before $5\times10^5$~s with a power-law to obtain a decay index of $\alpha_{1,X}= 1.4$--1.7 depending on whether we include the X-ray data before $7\times10^4$~s.
Although the X-ray decay index has a considerable range, even for the extreme case
of $\alpha_{1,X}= 1.7$, the extrapolation of the single power-law ($4\times10^{-15}$\flux) is inconsistent with \chandra\ upper limit of $2\times10^{-15}$\flux at  $3.4\times10^6$~s (Cenko et al.\ 2007), strongly suggesting the existence of a break
in the X-ray light curve.

We tested the standard afterglow/jet models (e.g., Sari et al. 1997; Rhoads 1999 and see Zhang \& \Mesz\ 2004 for a review) using our temporal decay indices and the spectral index of 
$\beta = 1.03\pm0.08$ that we derived from the X-ray data (consistent with the estimate from
 Burrows \& Racusin 2007b).
The pre-break spectral and temporal decay indices follow the relation $\alpha_1  = 3\beta/2$ $(1.58\pm0.12 \simeq 1.55\pm0.12)$, suggesting that the afterglow is at the evolutionary 
stage of $\nu_m < \nu < \nu_c$ where $\nu_m$ and $\nu_c$ are the typical frequency of the synchrotron radiation and the cooling frequency, respectively.
At this stage, the standard models for a constant ISM density predict 
a post-decay index $\alpha_2$ between
    $\alpha_2=\alpha_1+0.75 \simeq 2.33 \pm 0.12$ and
    $\alpha_2=2\beta + 1 \simeq 3.1 \pm 0.16$ depending
   on whether or not sideways expansion of the jet is included.
   Our post-decay index of $\alpha_2=2.87\pm0.17$ is closest to the
   value expected with maximal sideways expansion.

Thus, GRB~070125 shows a ``textbook jet break''.
Although the optical light curve provides much tighter constraints on the pre and post-break decay slopes, 
the break is independently seen in both optical and X-ray light curves,
and the consistency of the optical and X-ray light curves suggests that the break is achromatic.
Furthermore, the spectral and temporal decay indices before and after the break satisfy the predictions from the standard jet models.

\subsection{GRB~070311}
The late-time optical light curve of GRB~070311 contains a significant flare at $2\times10^5$~s, which is immediately followed by a steep decay 
during the period from $2.3\times10^5$ to $3.5\times10^5$~s with a decay slope of $3.1\pm0.4$ with $\chi^2/dof = 0.1/2$ (Figure~2 upper panel). 
A detailed analysis of this flare is provided by Guidorzi et al. (2007b). 
Before the flare, the four optical data points ($10^3$--$10^5$~s) set a decay slope of $\alpha_1 = 0.50\pm0.03$ ($\chi^2/dof = 4.4/2$).  After the flare, the two LBT data points are significantly below the extrapolation of the power-law decay before the flare, and have a decay slope of $\alpha_2 = 1.6\pm0.3$ ($\chi^2/dof = 0/0$).  
This indicates a break in the optical light curve,
with the break occurring during the flare from $10^5$~s to $3.5\times10^5$~s.
The post-break decay slope also suggests a relatively low electron index of $p\sim1.6$.
The XRT X-ray light curve also shows a flare very close to the optical flare, and it follows a similar steep post-flare decay slope ($3.3\pm0.5$).  
The X-ray and optical light curves after the flare and the break are mutually consistent.  
However, it is difficult to determine whether the decay slope for the last few X-ray data points has flattened to match our
LBT data given the S/N of the X-ray light curve.
It is also obvious that the optical-to-X-ray flux ratio has changed before and after the flare/break, suggesting that the early X-ray light curve has a significant flux contribution from additional emission components.

\subsection{GRB~070411}
GRB~070411 (Figure~2 lower panel) at $z=2.954$ (Jakobsson et al.\ 2007) has a well-sampled optical light curve at early times (before $6000$~s)
 and the early light curve shows complicated decay patterns that are also 
present in many other bursts (e.g., GRB~060206, Stanek et al.\ 2007).  
The late time optical light curve (after $10^4$~s), however, is not well 
sampled.  
We fit the optical light curve after $1000$~s with a single power-law and 
obtained a decay index of $\alpha = 0.94\pm0.01$ ($\chi^2/dof = 119.6/21$).  
Although this power-law fit cannot 
fully represent the complicated early afterglow decay, it is consistent with the late time optical light curve, where fitting the data after $4500$~s yields $\alpha = 0.94\pm0.02$ with improved fits with $\chi^2/dof = 13.4/6$.
However, our 3$\sigma$ upper limit from LBT at 10.3 days is slightly below the 
extrapolation of the single power-law determined from the early epochs of 
optical data.  
We also tested fitting the power-law decay to a longer time baseline that included the early optical data before $1000$~s, and we obtained slightly flatter slopes which will lead to larger deviations from the LBT limit at 10.3 days.  
The X-ray light curve has a slightly steeper slope for the single power-law fit ($\alpha_X=1.02\pm0.02$ with $\chi^2/dof = 19.5/9$) than the optical slope.
Since the last two X-ray data points show possible deviations from the single power-law decay, we also tried a broken power-law fit, and obtained a comparable fit with $\chi^2/dof = 17.9/7$).  
The optical and X-ray data set a lower limit for the jet break $t_b > 2.8\times10^5$~s.
A break between $3\times10^5$~s and $10^6$~s is consistent with the optical and X-ray data; however, the late time data are still not deep enough to exclude the single power-law model.

\subsection{GRB~070412, GRB~070419A, and GRB~070518}
GRB~070412 appears optically dark.  We observed GRB~070412 twice with the LBT at 2.8 hours and 8.2 days after the burst respectively. 
We place upper $3\sigma$ limits of $r'>25.2$ for both observations.  Most of the other optical observations of GRB~070412 also only yield upper limits.  Therefore, we are unable to search for the jet break in GRB~070412.
The late time optical light curves for both GRB~070419A and GRB~070518 flattened compared 
to the early optical observations taken before the LBT data,
indicating other emission components besides the afterglow.
In GRB~070518, the extra optical flux contribution may come from the host 
galaxy because the LBT images show that the source is extended.
GRB~070419A is more interesting, since there is a late time bump in the optical light curve, consistent with a supernova bump.  
The details will be discussed in companion papers (Garnavich et al.\ 2008, in preparation; McClelland et al. 2008 in preparation).
In both cases, we are unable to search for a late time optical jet break.

\section{Discussion\label{sec:dis}}
We observed six GRBs afterglows using the 8.4m LBT from 2.8 hours to 30.8 days after the burst triggers. 
With 20--30 minute exposures, we are able to reach a flux limit of $\sim25$--$26$~mag in the Sloan $r'$ band.
These observations systematically probe the late time decay of the afterglows.  
In particular, these observations enable us to extend the temporal baseline for 
measuring the jet breaks by another decade in a logarithmic time scale.  

In three of the six bursts, we are unable to detected the optical jet break because of the nature of the afterglow.  GRB~070412 is optically dark  
and the late-time optical light curves of GRB~070419A and GRB~070518 contain 
other emission components from the host galaxy or a supernova which will 
mask a jet break.
In the other three cases, we clearly detected jet breaks in GRB~070125
 and GRB~070311, and a break is probably required for GRB~070411.
The optical jet breaks in all three cases occur at late-times, when the 
afterglows have faded to be fainter than $R=24$~mag.
Such faint breaks are not detectable without deep, late-time imaging of the afterglow.
This strongly suggests that most failures to detect jet breaks
are due to the lack of well-sampled afterglows either in the rate of sampling or the temporal baselines.
Several other well-sampled \swift\ bursts (e.g., GRB~060206, Stanek et al.\ 2007 and GRB~060526, Dai et al.\ 2007) also clearly show jet breaks.

The XRT light curves suffer from similar problems.  Although the XRT light curves have good S/N at early times, the S/N decreases significantly for the time
scales on which the jet breaks frequently occur.  
The problem is particularly severe for the last few XRT data points, where the count rate is affected by the choice of source region, background region, binning, and other effects such as flares either in the GRB itself or in background sources.
Thus, conclusions about the existence of jet breaks with typical data 
are heavily dependent on the 
prior for fitting either a single power-law or broken power-law.
This is also reflected in our sample.  For example, in GRB~070125, the XRT data points can be fit with both single or broken power-laws.  
This problem is also present in several other bursts (e.g., GRB~060526, Dai et al.\ 2007), where the optical data 
require a break while the X-ray data are only consistent with a break.
In the case of GRB~070125, an upper limit from a deep \chandra\ observation (Cenko et al. 2007)
is below the extrapolation from the single power-law decay, independently confirming the existence of 
of a break.

Analyses of jet breaks are also affected by the complicated decay patterns.  
When extra emission processes, such as energy injection or flares, are important in the light curve, it is difficult to determine the forward shock emission decay slope.  
This may explain the case of GRB~070311, where extra emission components in both the optical and X-ray light curves may affect the slopes of the afterglows.

Given these problems, and the fact that almost all well-sampled bursts in the 
optical bands show a significant break, we argue that the basic picture that GRBs are collimated is still valid.
Invalidating the jet model requires examples without jet breaks even at late 
times rather than a failure to find a break at early times.
More well-sampled bursts are still needed to securely measure a large sample of jet breaks, to constrain the jet models and determine the beaming correction for GRBs.  
Accurate estimates
          of the beaming corrections are needed to determine the 
intrinsic rate of GRBs and to match those rates to possible progenitor
populations.

\acknowledgements 
We would like to thank D. Hartmann, A. Updike, B. Zhang, and N. Gehrels for their comments.
We thank the \swift\ team for the prompt detection and localization of GRBs and the rapid release of data products, the GRB Coordinates Network (GCN) and astronomers who contribute to the GCN circulars.
We thank everyone who contributed to the LBT/LBC Science Demonstration Time observations.

\clearpage

\begin{deluxetable}{llcccc}
\tabletypesize{\scriptsize}
\tablecolumns{6}
\tablewidth{0pt}
\tablecaption{LBT Photometry of GRBs \label{tab:data}}
\tablehead{
\colhead{Burst} &
\colhead{Observation} &
\colhead{Days} &
\colhead{Filter} &
\colhead{Exposures} &
\colhead{Mag} \\
\colhead{} &
\colhead{Time} &
\colhead{After} &
\colhead{} &
\colhead{(sec)} &
\colhead{} \\ 
\colhead{} &
\colhead{(UT)} &
\colhead{Trigger} &
\colhead{} &
\colhead{} &
\colhead{} 
}
\startdata
GRB~070125  & 2007-02-21.10  & \phn26.80 & $r'$ &  $10\times200$ & $26.3\pm0.3$ \\
            & 2007-11-19.50  & 298.2\phn   & $r'$ &  $\phn9\times200$  & $\phm{00.00}>26.2$ \\
GRB~070311  & 2007-03-17.14  & \phn\phn6.13 & $r'$ &  $10\times200$ & $24.73\pm0.06$ \\
            & 2007-03-20.13  & \phn\phn9.05 & $r'$ &  $15\times200$ & $25.42\pm0.12$ \\
GRB~070411  & 2007-04-15.15  & \phn\phn3.30 & $r'$ &  $10\times200$ & $23.95\pm0.05$ \\
            & 2007-04-22.14  & \phn10.30 & $r'$ &  $10\times200$ & $\phm{00.00}>25.3$ \\
GRB~070412  & 2007-04-12.18  & \phn\phn0.12 & $r'$ &  $10\times200$ & $\phm{00.00}>25.2$ \\
            & 2007-04-20.31  & \phn\phn8.20  & $r'$ &  $\phn5\times200$  & $\phm{00.00}>25.2$ \\
GRB~070419A & 2007-04-23.15  & \phn\phn3.70  & $r'$ &  $10\times200$ & $24.75\pm0.18$ \\
            & 2007-05-10.18  & \phn20.80 & $r'$ &  $25\times200$ & $25.29\pm0.05$ \\
            & 2007-05-20.22  & \phn30.80 & $r'$ &  $15\times200$ & $25.71\pm0.13$ \\
GRB~070518  & 2007-05-19.32  & \phn\phn0.72 & $r'$ &  $\phn5\times200$  & $23.03\pm0.05$ \\
            & 2007-05-20.31  & \phn\phn1.71 & $r'$ &  $\phn5\times200$  & $23.36\pm0.05$ \\
            & 2007-05-22.21  & \phn\phn3.60 & $r'$ &  $15\times200$ & $23.56\pm0.05$ \\
\enddata
\end{deluxetable}

\clearpage

\begin{figure}
\epsscale{1}
\plotone{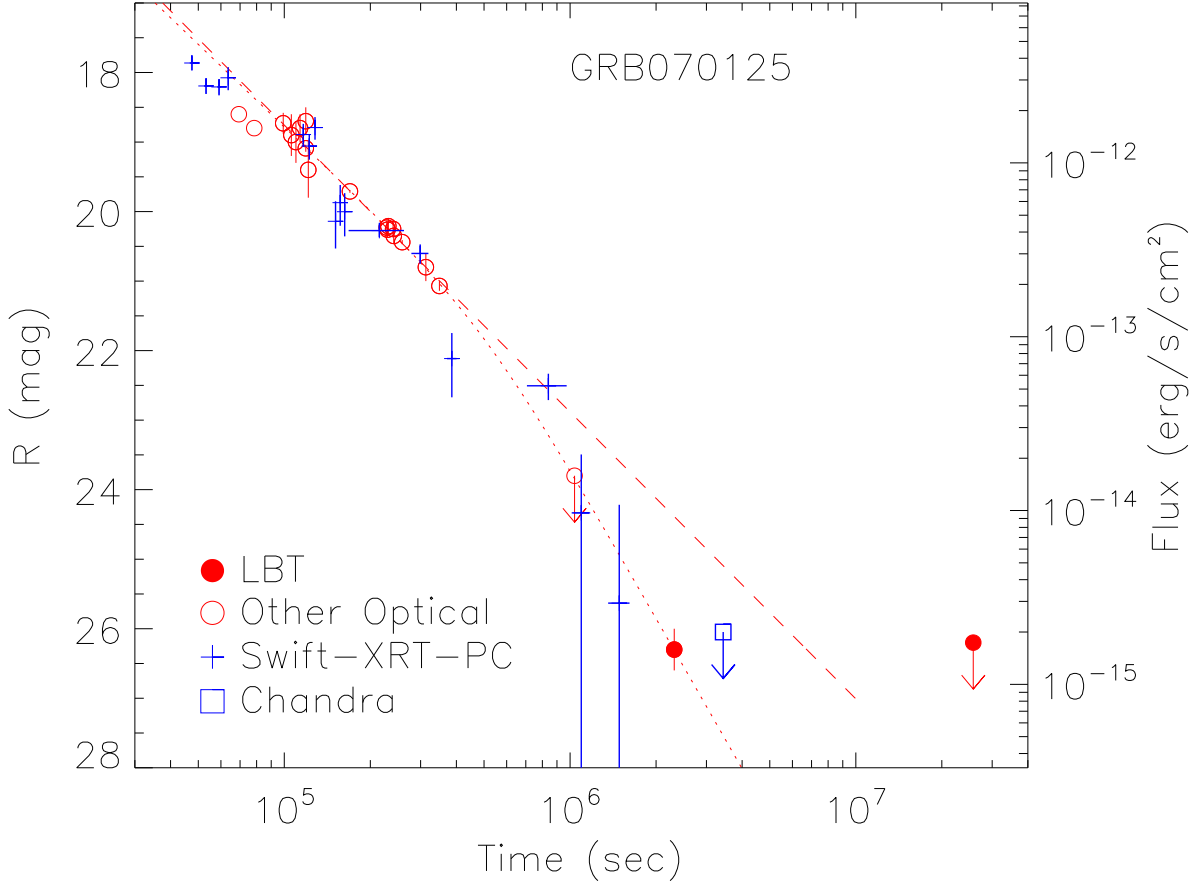}
\caption{Optical and X-ray light curves of GRB~070125.  
We included additional $R$ band data from the GCN circulars (Cenko \& Fox 2007a; Xing et al. 2007; Uemura et al. 2007; Haislip et al. 2007; Greco et al. 2007a; Yoshida et al. 2007a; Terra et al. 2007a; Mirabal et al. 2007).
The dashed line is a single power-law fit ($\alpha=1.65\pm0.05$) to the optical data from $9\times10^4$ to $4\times10^5$~s.
The dotted line is a broken power-law fit to the whole optical data set with $\alpha_1=1.58\pm0.12$, $\alpha_2=2.87\pm0.17$, and $T_b = (5.0\pm1.6)\times10^5$~s.
The X-ray and optical data are aligned such that the steepest single power fit to the X-ray data ($\alpha_X = 1.7$) overlaps the single power fit to the optical data (the dashed line).
\label{fig:lcone}}
\end{figure}

\begin{figure}
\epsscale{0.7}
\plotone{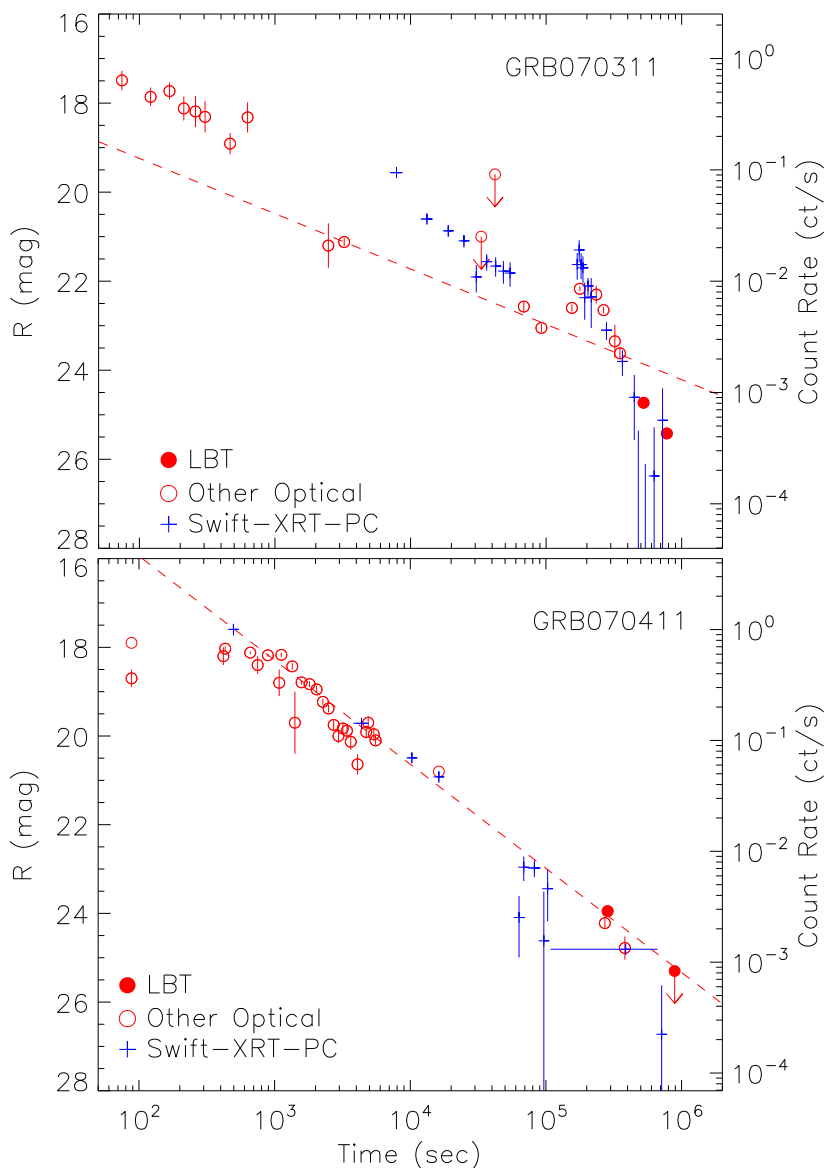}
\caption{Optical and X-ray light curves of GRB~070311 and GRB~070411.
We included additional $R$ band data from Guidorzi et al. (2007) and GCN circulars (Yoshida et al. 2007b; Halpern \& Armstrong 2007abcd; Jelinek et al. 2007a; Updike et al. 2007a; Greco et al. 2007b; Kann et al. 2007a) for GRB~070311, and for GRB~070411 (Mikuz et al. 2007a; Gomboc et al. 2007; Jelinek et al. 2007b; Berger et al. 2007; Mikuz et al. 2007b; Kann et al. 2007b; Ferrero et al. 2007; Perley et al. 2007).
\label{fig:lctwo}}
\end{figure}

\end{document}